\documentclass[aps,prb,reprint,groupedaddress,showpacs]{revtex4-1}

\usepackage{graphicx}
\usepackage{dcolumn}
\usepackage{bm}
\usepackage{color}
\usepackage{amsmath}

\begin{document}


\title{Tuning of the electrically evaluated electron  Land\'{e} $g$-factor in GaAs quantum dots and quantum wells of different well widths}


\author{G. Allison$^{1,2,3}$}
 \email[]{giles.allison@riken.jp, tarucha@ap.t.u-tokyo.ac.jp}
\author{T. Fujita$^{1}$}
\author{K. Morimoto$^{1}$}
\author{S. Teraoka$^{1}$}
\author{M. Larsson$^{1}$}
\author{H. Kiyama$^{1,4}$}
\author{A. Oiwa$^{1,4}$}
\author{S. Haffouz$^{5}$}
\author{D. G. Austing$^{5}$}
\author{A. Ludwig$^{6}$}
\author{A. D. Wieck$^{6}$}
\author{S. Tarucha$^{1,3}$}

\affiliation{$^1$Department of Applied Physics, The University of Tokyo, 7-3-1 Hongo, Bunkyo-ku, Tokyo 113-8656, Japan}
\affiliation{$^2$Department of Physics, Princeton University, Princeton, NJ 08544, USA}
\affiliation{$^3$Center for Emergent Matter Science (CEMS), RIKEN, 2-1 Hirosawa, Wako-shi, Saitama 351-0198, Japan}
\affiliation{$^4$Institute of Scientific and Industrial Research, Osaka University, 8-1 Mihogaoka, Ibaraki-shi, Osaka 567-0047, Japan}
\affiliation{$^5$National Research Council of Canada, M50, Montreal Rd, Ottawa, Ontario K1A 0R6, Canada}
\affiliation{$^6$Lehrstuhl f\"{u}r Angewandte Festk\"{o}rperphysik, Ruhr-Universit\"{a}t Bochum, Universit\"{a}tsstra{\ss}e 150, Geb\"{a}ude NB, D-44780 Bochum, Germany}


\date{October 14, 2014}

\begin{abstract}
We evaluate the Land\'{e} $g$-factor of electrons in quantum dots (QDs) fabricated from GaAs quantum well (QW) structures of different well width. We first determine the Land\'{e} electron $g$-factor of the QWs through resistive detection of electron spin resonance and compare it to the enhanced electron $g$-factor determined from analysis of the magneto-transport. Next, we form laterally defined quantum dots using these quantum wells and extract the electron $g$-factor from analysis of the co-tunneling and Kondo effect within the quantum dots. We conclude that the Land\'{e} electron $g$-factor of the quantum dot is primarily governed by the electron $g$-factor of the quantum well suggesting that well width is an ideal design parameter for $g$-factor engineering QDs.
\end{abstract}


\maketitle


\section{Introduction}

The Zeeman splitting of energy levels with different spin in a magnetic field is an important parameter in the growing field of spin-based quantum computation. The ability to engineer the Land\'{e} $g$-factor of carriers in semiconductor structures is an important step in spintronics in which the different spins must be individually addressed. For example, quantum information processing based on quantum dot spin qubits requires accurate, coherent, and selective control of the rotation of single electron spins. The rotations may be accurately controlled with time-dependent magnetic fields by conventional electron spin resonance measurements, however, selectivity requires unique resonant frequencies for all qubits within the array or the ability to control the time-dependent local magnetic field. One scheme to achieve this selectivity requires the ability to control the electron $g$-factor or modulate the $g$-tensor \cite{kato, pingenot1, pingenot2, deacon}. 

A further example of the importance of $g$-factor engineering is given in a proposal to transfer a coherent superposition of photon polarization states to a coherent superposition of electron spin states \cite{vrijen}. The proposal requires a V-shaped three-level system with small Zeeman energy of the conducting electrons compared with light hole Zeeman energy and photon bandwidth, i.e. $g_{e}\mu_{B}B\ll \Delta E_{ph} \ll g_{lh}\mu_{B}B$, where $g_{e}$ and $g_{lh}$ are the $g$-factors of electrons and light holes, respectively, $\mu_{B}$ the Bohr magneton, and $\Delta E_{ph}$ the photon bandwidth. This scheme has been experimentally demonstrated for an ensemble of photons and electrons in $g$-factor engineered QWs \cite{kosaka, kosaka2}. To enable practical quantum communication, coherent angular momentum transfer is required between a single photon polarization state and a single electron spin state. Such an interface would be the basis of a large scale quantum information network. It is for this reason that we investigate whether the electron $g$-factors in a QD in a QW structure can also be controlled by the QW width.

It is known that the $g$-factor of QDs formed in single GaAs/AlGaAs heterostructures (SH) does not differ significantly from the $g$-factor of the two dimensional electron gas (2DEG), which in turn is little different from the bare $g$-factor of bulk GaAs $|g|=0.44$ (for example Nowack \textit{et al.} \cite{nowack} found $|g|=0.39$). However, this value may be modified due to changes of the confining potential. In contrast, the $g$-factor of QDs fabricated in double AlGaAs/GaAs/AlGaAs heterostructures (DH) has not been reported yet, so it remains unknown how greatly they differ from the $g$-factor of the 2DEG and whether they are capable of satisfying the above requirement for small Zeeman energy of single electron spins for coherent transfer \cite{vrijen}.

In this work we report our experiments to engineer the electron Land\'{e} $g$-factor of QWs and subsequently QD systems based on AlGaAs/GaAs/AlGaAs DH by tuning the well width, $w$. In Refs. \citenum{lejeune, malinowski, dios, yugova, salazar} the $g$-factor is shown to be strongly influenced by the quantum confinement in the GaAs QW and has an anisotropy with respect to the magnetic field direction. The origin of the well width dependence of the $g$-factor is the penetration of the electron wavefunction into the AlGaAs barrier layer. It has been found that the $g$-factor of QDs in the QW substrate can be engineered by the well width because the lateral confinement does not strongly affect the $g$-factor.

Several techniques can be utilized to determine the electron $g$-factor such as the change in the magneto-resistivity induced by (electron spin resonance) ESR that has been used for GaAs/AlGaAs SHs \cite{dobers, teraoka}. Here we employ the same method to determine the well width dependence of the $g$-factor of QWs based on DHs. For the QDs we deduce the $g$-factor from the magnetic field dependence of electron transport from observations of co-tunnelling and the Kondo effect.

\begin{figure}[t]
\begin{center}
\includegraphics[width= 0.48\textwidth, keepaspectratio]{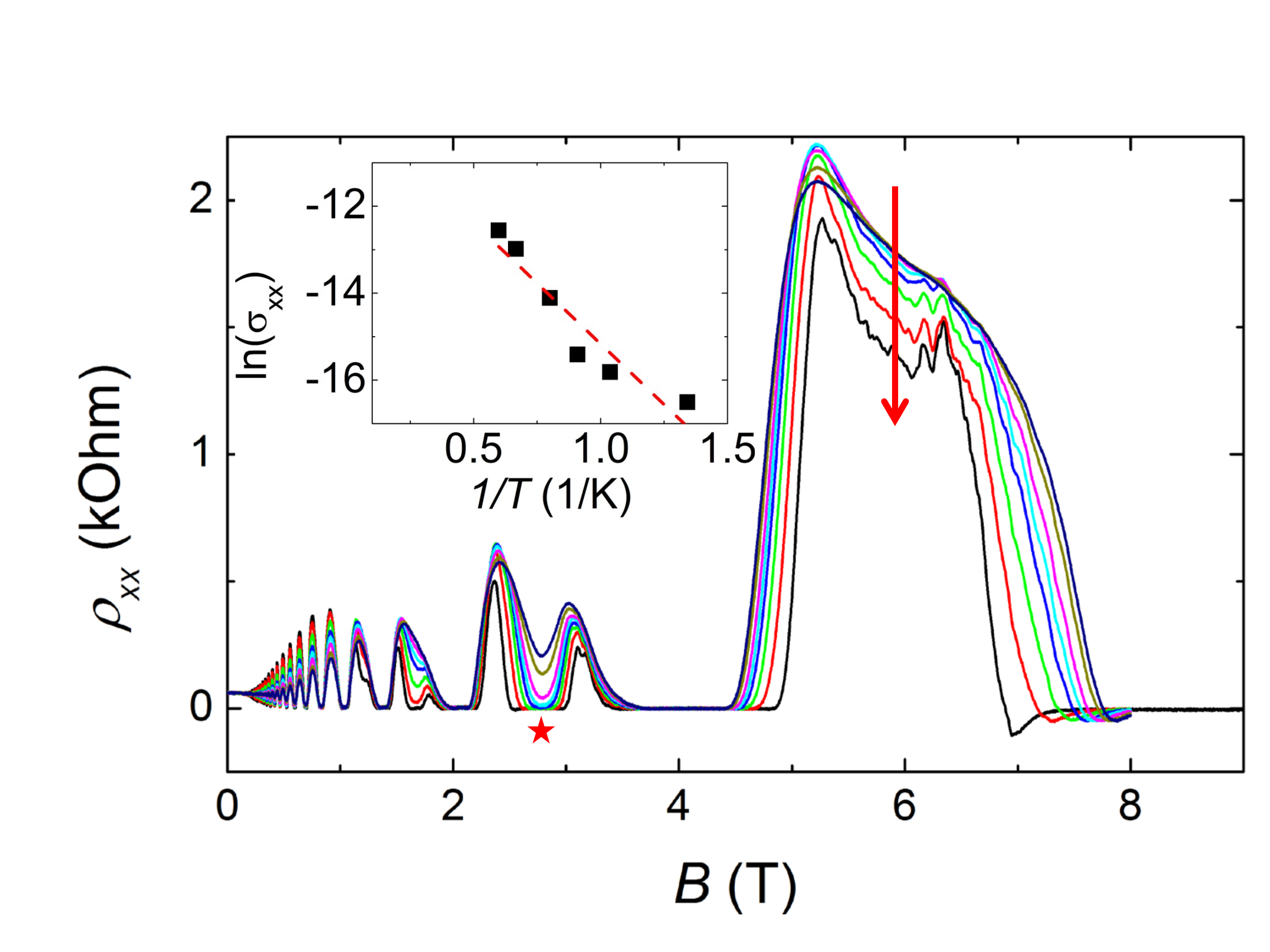}
\caption{Main: Magnetic field dependence of the longitudinal resistance of sample J65 for $T=$ 0.25-1.67 K. The arrow indicates the direction of decreasing temperature. Inset: Log plot of the conductance as a function of temperature at filling factor three as indicated by the asterisk in the main figure at $B=$ 2.8 T. The dashed line is the linear fit described in the main text.}
\label{magnetotransport}
\end{center}
\end{figure}

\begin{table}
\caption{QW characteristics.}
\begin{tabular}{lrrrrr}
\hline
\hline
Sample & J65 & J67 & J107 & 14155 & 14367 \\ 
\hline
\hline
QW width $w$ (nm) & 7.0 & 9.2 & 13.0 &  7.3 & 9.4-14.4 \\
\hline
$n_{2D} (\times 10^{11}$ cm$^{-2}$) & 2.10 & 2.30 & 3.75 & 2.1 & 2.5 \\ 
\hline
$\mu (\times 10^{6}$ cm$^{2}/$Vs)  & 0.50 & 0.37 & - & 0.1 & 0.3-0.5 \\ 
\hline
Bare Land\'{e} $g$-factor  & 0.15 & 0.22 &  0.39 & 0.12 & - \\
\hline
\end{tabular}
\label{tablesummary}
\end{table}

\section{Sample details \& characterisation}

The QWs were grown by molecular beam epitaxy and have the following growth sequence: GaAs substrate, 67 period GaAs/AlGaAs superlattice structure, 830 nm thick undoped GaAs layer, Al$_{x}$Ga$_{1-x}$As layer of thickness $d$, GaAs QW of thickness $w$, 30 nm thick undoped Al$_{x}$Ga$_{1-x}$As layer, 65 nm thick Si-doped Al$_{x}$Ga$_{1-x}$As layer, 5 nm thick GaAs capping layer. The quantum well widths of all the samples used in this work are summarized in Table \ref{tablesummary}. All samples have Aluminium content $x=0.265$, AlGaAs barrier thickness $d=20$ nm and Si-doping of 2 x $10^{18}$ cm$^{-3}$ except 14155 for which $x=0.34$, $d=100$ nm and Si-doping of 1 x $10^{18}$ cm$^{-3}$ and 14367  for which $x=0.3$, $d=100$ nm and Si-doping of 1 x $10^{18}$ cm$^{-3}$. Sample 14367 was grown with a gradient QW width by stopping the usual homogenising sample rotation during MBE-growth which yields a higher growth rate the closer the sample area is to the off-normal oriented effusion cells.

Samples were fabricated into a Hall bar geometry of size 300 $\mu$m x 90 $\mu$m. Additional Ti/Au surface gates were added and used to locally deplete electrons to form gate confined lateral quantum dots. The QD structures were optimized after carefully designing the surface gate geometry by performing an electrostatic potential calculation \cite{takakura} based on the carrier density and the distance from the surface gate electrodes to the two-dimensional electron gas.

The samples were placed in a short-circuited waveguide which forms a rectangular microwave cavity in a variable temperature $^{3}$He cryostat with the sample located at the shorted end of the cavity where the maximum of the magnetic field of a standing microwave can be applied \cite{teraoka}. First the resistance of the QWs was measured in a magnetic field parallel to the growth direction using a low frequency four-terminal lock-in technique and the electron concentration and mobility were determined at $T=250$ mK using the Fourier transform and applying the Drude model as summarised in Table \ref{tablesummary}.

\begin{figure}[t]
\begin{center}
\includegraphics[width= 0.48\textwidth, keepaspectratio]{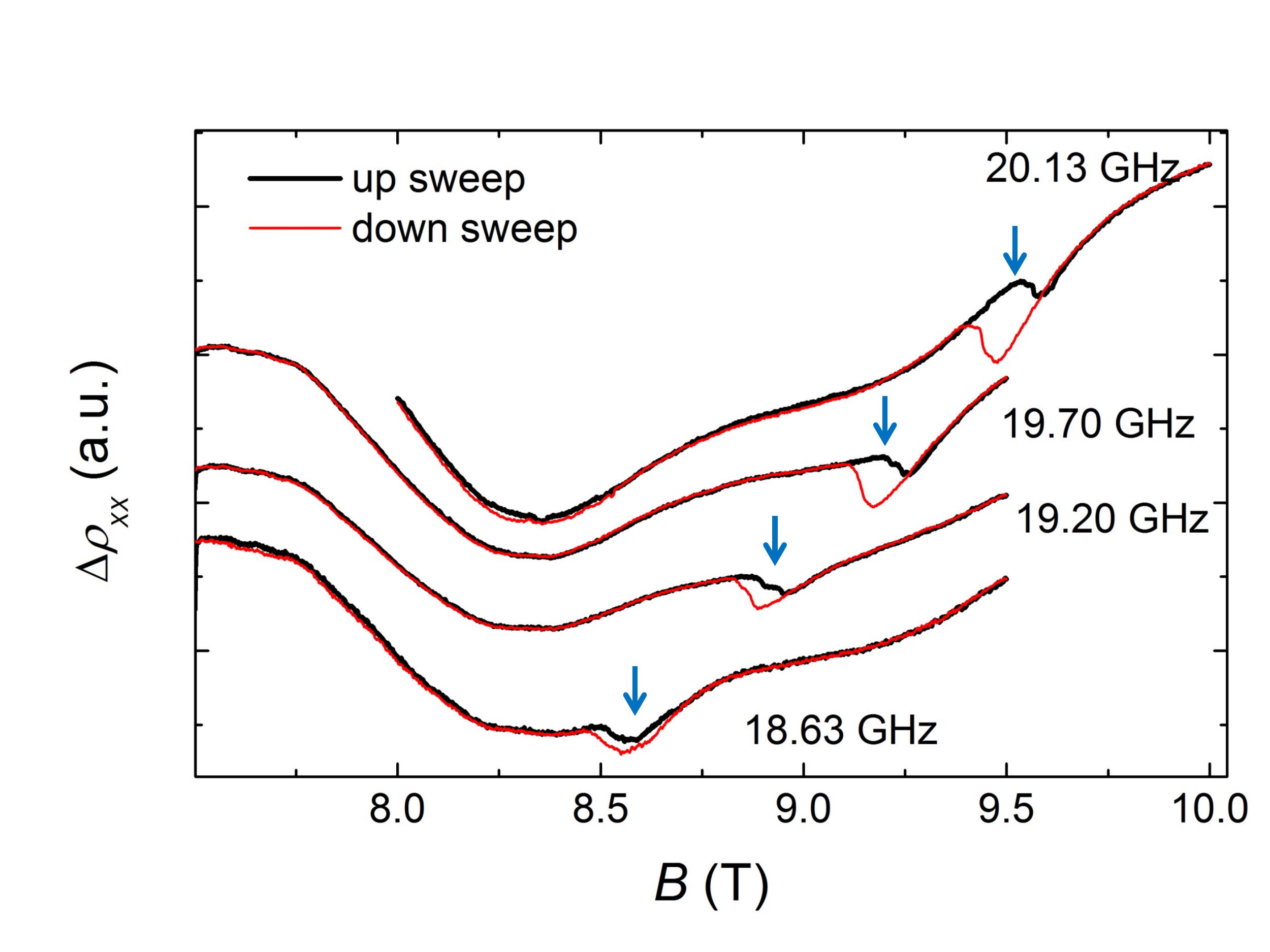}
\caption{Difference in longitudinal resistance with and without microwave radiation at different frequencies for both sweep directions of sample J65. Arrows indicate the resonant condition on the up sweep.}
\label{esrA}
\end{center}
\end{figure}

\begin{figure}[t]
\begin{center}
\includegraphics[width= 0.4\textwidth, keepaspectratio]{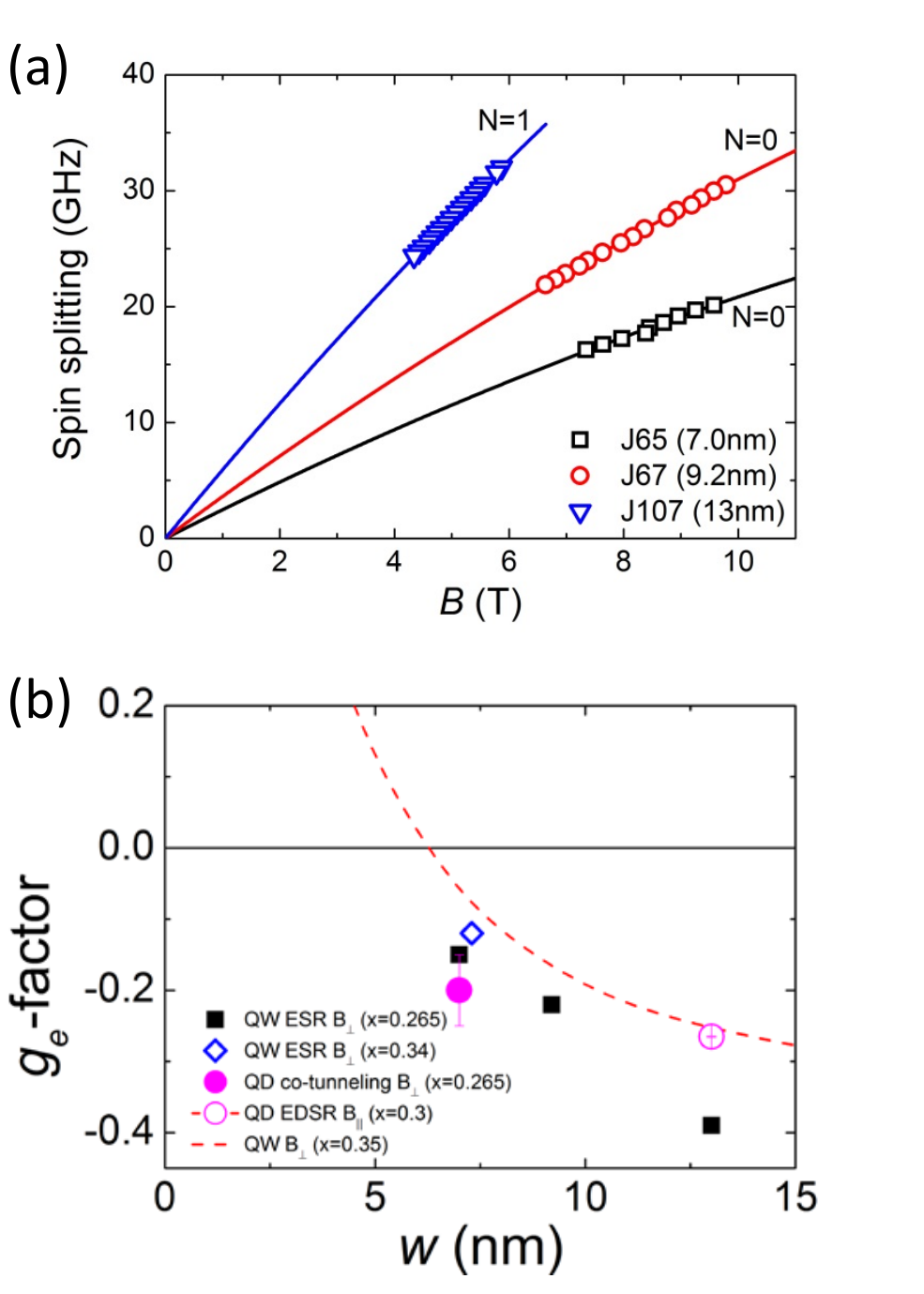}
\caption{(a) Magnetic field dependence of the spin splitting for different QWs and Landau level index. The fit explained in the main text is used to determine the bare electron $g$-factor. (b) Well width dependence of the electron g-factor of QWs and QDs, where $x$ is the aluminium content in the barrier. The dashed line is the calculated dependence of a QW according to Ref.\citenum{ivchenko}.}
\label{esrB}
\end{center}
\end{figure}

We next study in detail the temperature dependence of the spin splitting of the Landau levels, $\Delta E$, in the magnetotransport measurements. Following the procedure of Refs. \citenum{nicholas} and \citenum{usher} the temperature dependence of the magnetoresistance is shown for sample J65 in Fig. \ref{magnetotransport}. $\Delta E$ is determined by fitting the conductivity at odd integer filling factors (for example the spin-splitted third filling factor at $B=2.8$ T indicated by the asterisk at the local minima in the resistivity) with
\begin{equation}\label{xxx}
  \sigma_{xx}=\sigma_{0}\exp\left[-\frac{\Delta E}{2k_{B}T}\right],
\end{equation}
where $k_{B}$ is the Boltzman constant, $\Delta E = g^{*}\mu_{B}B$ and $g^{*}$ is the so-called exchange enhanced electron $g$-factor, which is known to be greatly enhanced from its bare value \cite{nicholas, usher}. At filling factor 3, we found $|g^{*}|=7.8$ for sample J65 and approximately $|g^{*}|=5$ for sample J67. These values of $g^{*}$ are indeed much greater than the bulk value of GaAs (-0.44) and, as will be shown, the bare electron $g$-factor of the QWs. Through careful analysis of the temperature dependence \cite{fletcher} we could also deduce the effective mass at low magnetic field for sample J65 $m^{*}=0.058m_{0}$ a 10\% reduction from the value of bulk GaAs $m^{*}=0.067m_{0}$. This reduction is proposed to be due to electron-electron interactions and is in agreement with quantum wells of similar electron concentration\cite{tan, hatke}.

\section{$g$-factor of quantum wells}

Next, we perform ESR measurements to extract the magnitude of the bare $g$-factor, $|g_{e}|$, unaffected by the exchange interaction.  In order to detect small changes in the resistivity $\rho_{xx}$ along the direction of current flow we employ a double lock-in technique \cite{teraoka}. The resistivity of the sample at odd filling factors is measured as a function of magnetic field perpendicular to the QW plane both with and without a range of applied microwaves. The difference in longitudenal resistance with and without microwave, $\Delta \rho_{xx}$, for sample J65 at filling factor 1 is shown in Fig. \ref{esrA}. Clear resonances are seen, which go to higher magnetic fields as the microwave frequency is increased. When the condition $hf=g_{e} \mu_{B}B$, where $f$ is the microwave frequency, and $h$ the Planck constant, is satisfied, the microwave may flip the spin of conducting electrons, which results in scattering of the carriers between edge channels and this is observed as a change in the resistivity. On resonance, neighbouring nuclear spins become polarized through the hyperfine interaction. The resulting effective nuclear magnetic field acts to shift the resonance to lower magnetic field on the down sweep causing a hysteresis in the magnetoresistance known as an Overhauser shift\cite{teraoka, berg}. It should be noted that the hysteresis for our samples is surprisingly small in comparison to Refs. \citenum{teraoka} and \citenum{berg}. The magnetic field dependence of the resonant frequency is fitted with\cite{dobers}
\begin{equation}\label{yyy}
   hf=\left[g_{e}-c\left(N+\frac{1}{2}\right)B\right]\mu_{B}B,
\end{equation}
where $c$ is a fitting parameter and $N$ is the Landau level index as shown in Fig. \ref{esrB}(a).

The bare electron $g$-factor was determined for four quantum wells and a clear dependence on the well width is seen in Fig. \ref{esrB}(b). Although it is not possible to determine the sign of the $g$-factor using this method, through comparison with theory and previous experiments results it is assumed to be negative for the range of well widths studied. The data are in qualitative agreement with experiments on GaAs multiple quantum wells determined using time resolved photoluminescence \cite{lejeune} and the corresponding calculation using a one-band approximation\cite{ivchenko}. The lack of complete coincidence between our results and previous work is due to the differences in Al content of the AlGaAs barrier (indeed the $g$-factor of our sample 14155 with $x=0.34$ is much closer to the previous work with $x=0.35$ than our other samples with $x=0.265$). The bare electron $g$-factor for sample J65 is a factor of 20 smaller than the exchange enhanced $g$-factor determined from Eq. \ref{xxx}, which is in qualitative agreement with results for a GaAs modulation-doped GaAs/n-AlGaAs single heterostructure \cite{usher}.

\begin{figure}[t]
\begin{center}
\includegraphics[width= 0.48\textwidth, keepaspectratio]{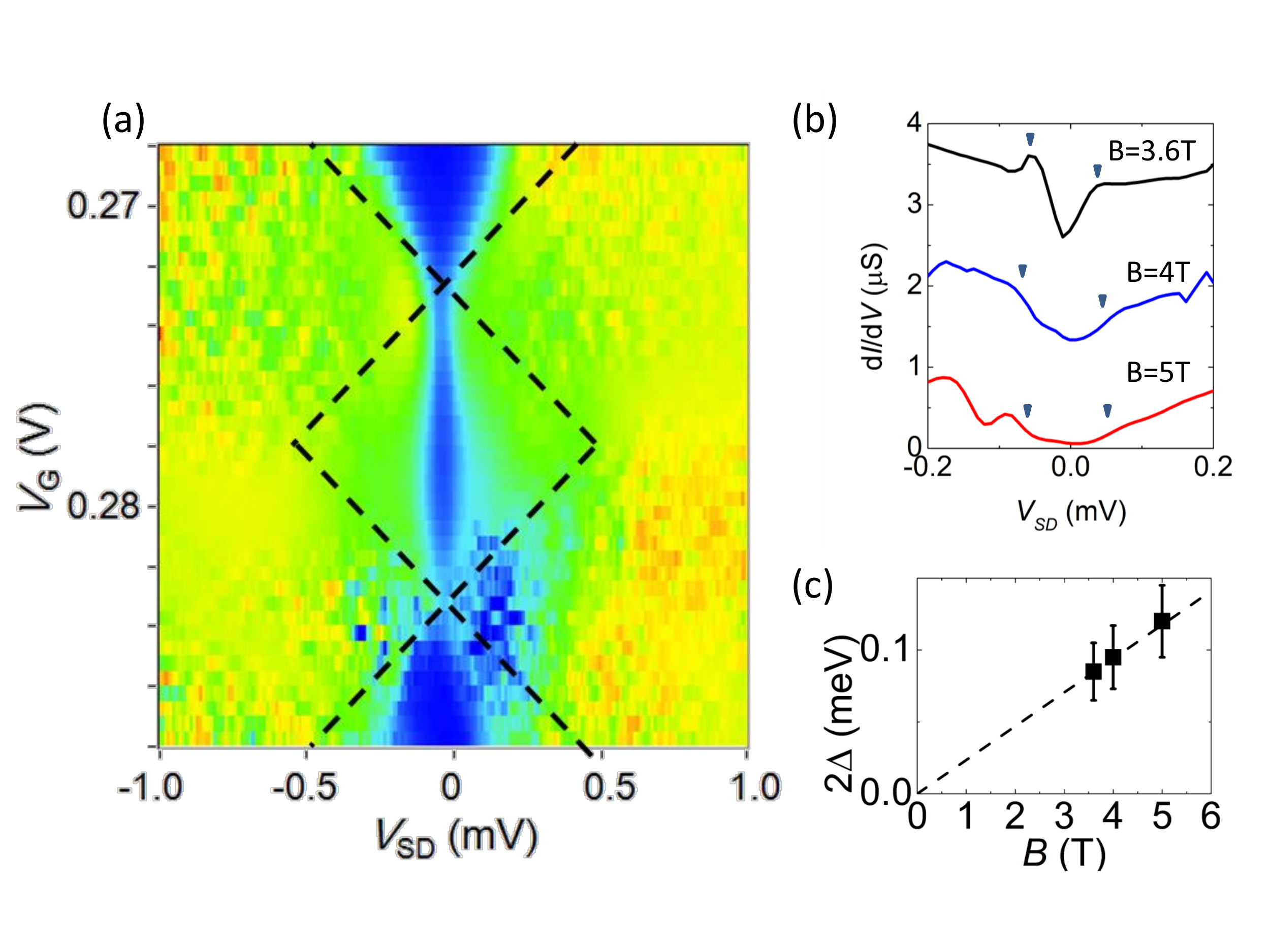}
\caption{Extracting the $g$-factor of sample J65 from co-tunneling events. (a) Numerical derivative $dI_{SD}/dV_{SD}$ in arbitrary units showing a typical Coulomb diamond with odd electron number under a perpendicular magnetic field $B=$ 4 T, where the inelastic cotunneling effect is observed. (b) 1D traces of differential conductance taken at the centre of the Coulomb diamond at $B=$ 3.6, 4, and 5 T. As the magnetic field is increased, the spacing of the step structure monotonically increases. (c) The splitting energies extracted from (b) as a function of magnetic field. The linearity of the increase implies that the inelastic cotunneing effect arises from the Zeeman splitting of the degenerate energy levels.}
\label{cotunneling}
\end{center}
\end{figure}

\section{$g$-factor of quantum dots}

In previous studies it was found that the $g$-factor of a QD fabricated in a single heterostructure is shifted only slightly from the value for the 2DEG system, implying that the effect of the confinement due to the lateral electrostatic gate potential might be weak. The effect of the confinement potential may also be weak for QDs fabricated in a QW system, however, this has not been examined yet. In this section we confirm this to be the case and that QDs fabricated from our QWs also have small electron $g$-factor.

Next, we confirm that the electron $g$-factor in QDs fabricated from the QWs is consistent with the ESR measurements. The following experiments were performed in a $^{3}$He/$^{4}$He dilution refrigerator at a bath temperature of $T=$ 140 mK. Typically the electron $g$-factor of a QD is determined from excited state spectroscopy, however, this method is hard to realise in our devices due to the small energy splitting at small g-factor. Instead we use the various methods as sensitive sensors for the $g$-factor; inelastic co-tunneling, the Kondo effect for single QDs, and EDSR with Pauli spin blockade for a double QD.

Systematic study of the $g$-factor evaluation methods in a few electron quantum dots is reported in Refs. \citenum{Zumbuhl, Kogan}. We form a gate defined lateral QD from sample J65 and determine the numerical derivative of the current,  $I_{SD}$, in a magnetic field perpendicular to the QW plane (as shown in Fig. \ref{cotunneling}(a) for an odd electron number, $N=$ 25, at a magnetic field $B_{\perp}=$ 4T). Peak splitting features are observed around the zero-bias point that are due to inelastic spin-flip cotunneling events via the two Zeeman-split energy levels in the QD. Such a co-tunneling event can occur for an odd number of electrons when the source-drain bias exceeds the splitting of the Zeeman energy levels, i.e. $e|V_{SD}|>g\mu_{B}|B|$. As a result, step-like features are observed in the differential conductance for both positive and negative bias, and are separated by twice the Zeeman splitting energy yielding directly the $g$-factor. 

The cotunnelling curves at $B_{\perp}=3.6$, 4 and 5 T are plotted in Fig. \ref{cotunneling}(b), offset by 1 $\mu$S. The peaks become less pronounced at larger magnetic field. The energy splitting width is estimated at the marked point and plotted in Fig. \ref{cotunneling}(c) as a function of magnetic field. The splitting energy is proportional to the applied magnetic field, giving a strong evidence that the observed cotunnelling signals originate from inelastic spin-flip cotunneling. From linear fitting, the electron $g$-factor for the perpendicular magnetic field is estimated to be $|g_{e}^{\perp}|$ = 0.20 $\pm$ 0.05, which is in good agreement with the value of 0.18 deduced from the ESR measurements described above. This result suggests that the $g$-factor is predominantly governed by the width of the QW (of the order of 10 nm) such that the lateral confinement within the QW (of the order of 100 nm) plays little part.

\begin{figure}[t]
\begin{center}
\includegraphics[width= 0.48\textwidth, keepaspectratio]{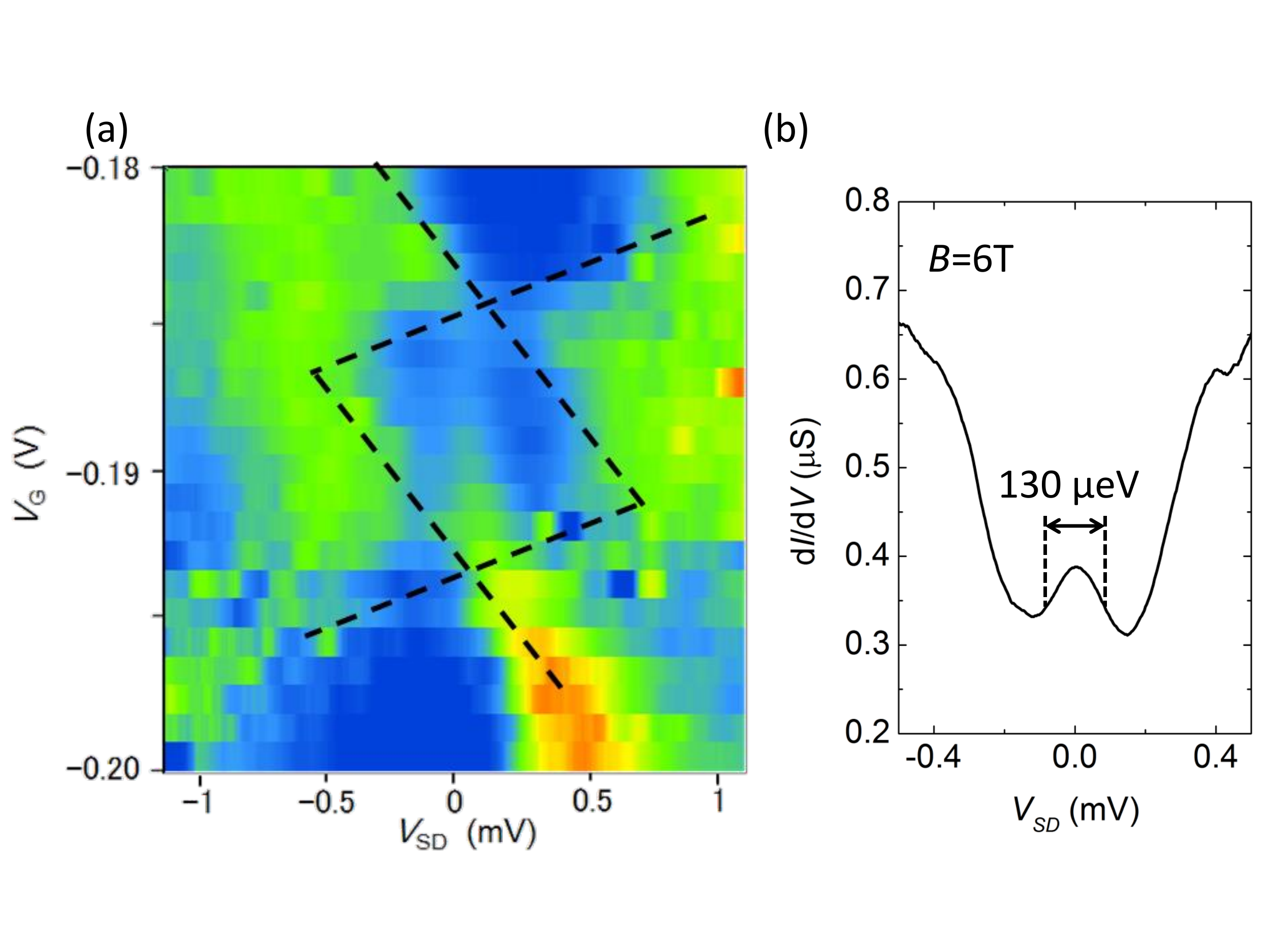}
\caption{Extracting the $g$-factor of sample J65 from the Kondo effect. (a) Numerical differential conductance of a QD under an applied in plane magnetic field of 6 T. Dashed lines are guide for the eye. (b) 1D trace of the differential conductance at fixed gate voltage showing no splitting of the zero-bias anomaly.}
\label{kondo}
\end{center}
\end{figure}

We then perform similar measurements on a different QD from the same wafer (J65) in a magnetic field parallel to the plane of the QW at $T=0.3$ K. Fig. \ref{kondo}(a) shows a Coulomb diamond of the QD at $B_{\parallel}=6$ T that is tilted in comparison to Fig. \ref{cotunneling} due to an asymmetry in the coupling strength between the QD and the leads. The Coulomb diamond is dominated around zero bias by an anomalous single peak due to the Kondo effect. In Ref. \citenum{kanai} the electron $g$-factor has been evaluated from the observed splitting of this Kondo zero-bias anomaly in a magnetic field, however, the spin-1/2 Kondo effect has never previously been observed in QDs in such large magnetic fields. Fig. \ref{kondo}(b) shows a 1D slice through the Coulomb diamond and it is clear that no splitting of the zero-bias anomaly is observed as a function of source and drain voltage, even at this large magnetic field. It is only possible to observe splitting at magnetic fields for which $2g\mu_{B}B$ exceeds the full width half maximum (FWHM) of the peak at zero magnetic field. For a FWHM of 130 $\mu$eV at $B_{\parallel}=6$ T this imposes an upper limit on the electron $g$-factor of $|g_{e}^{\parallel}|<$0.19. Similar experiments were performed in the perpendicular magnetic field orientation, which gave $|g_{e}^{\perp}|=0.2$. The results in both perpendicular and parallel magnetic fields are in good agreement with the results from the ESR measurements, however, it is not possible to observe the anisotropy of the electron $g$-factor reported in Ref. \citenum{lejeune} because only an upper bound of $g_{e}$ can be determined from the measurements.

Furthermore, we have determined the $g$-factor from electric dipole spin resonance (EDSR) measurements in an in-plane magnetic field of a double quantum dot formed from sample 14367 with a 13 nm thick QW where ESR-induced spin-flips are detected in the Pauli spin blockade regime \cite{Koppens}. We find  $|g_e^{||}|= 0.265$ for a well width $w = 13$nm with an Al content of the barrier $x=0.3$. This result agrees well with the bare QW $g$-factor in Fig. 3(b) as the relatively larger Al content in the barrier of the double quantum dot device should give a relatively more positive $g$-factor.

Finally, we consider the range of $g$-factor necessary for use in the V-shaped three-level system proposed for coherent angular momentum transfer between photons and electrons \cite{kosaka}. Typically, a Ti:Sapphire laser, in the pico second pulse mode used in our experiments, has a bandwidth of the order 0.6 meV, which is more than a factor of ten larger than the Zeeman energy for a quantum dot with $|g_{e}|=0.2$ at  a magnetic field of $B=6$ T. This suggests that a quantum well width between 4 and 10 nm is suitable.

\section{Conclusion}

In summary, we have observed a clear well width dependence of the electron Land\'{e} $g$-factor in GaAs DH quantum wells using resistively detected electron spin resonance techniques. The $g$-factor determined from the Kondo effect and co-tunnelling processes of a quantum dot based on a quantum well of 7 nm width indicates that the $g$-factor of a QD is mainly governed by the QW $g$-factor and thus the quantum well width. Electron $g$-factors sufficiently small to be suitable for use as part of a quantum repeater in a large scale quantum information network are experimentally achievable and designable by the QW parameters.

\bigskip
\begin{acknowledgments}
This work was supported by Grants-in-Aid for Scientific Research A (Grant No. 25246005), S (Grant No. 26220710), and Innovative Area “Nano Spin Conversion Science” (Grant No. 26103004), FIRST program, Intelligence Advanced Research Projects Activity project “Multi-Qubit Coherent Operations” through Copenhagen University, MEXT Project forDeveloping Innovation Systems, The Strategic Information and Communications R\&D Promotion Programme (SCOPE) of the Ministry of Internal Affairs and Communications Government of Japan (MIC), and ImPACT Program of Council for Science, Technology \& Innovation (Cabinet Office, Government of Japan). T.F. is supported by JSPS Research Fellowships for Young Scientists.M.L. acknowledges support as an “International Research Fellow of the Japan Society for the Promotion of Science”. A.L. and A.D.W. acknowledge gratefully support from Mercur Pr-2013-0001, BMBF - Q.com-H 16KIS0109, and DFH/UFA CDFA-05-06.
\end{acknowledgments}

\end{document}